\documentclass[conference]{IEEEtran}
\IEEEoverridecommandlockouts
\usepackage{cite}
\usepackage{graphicx}
\usepackage{listings}
\def\BibTeX{{\rm B\kern-.05em{\sc i\kern-.025em b}\kern-.08em
    T\kern-.1667em\lower.7ex\hbox{E}\kern-.125emX}}
\begin{document}

\title{Capability-based access control for multi-tenant 
systems using OAuth 2.0 and Verifiable Credentials}

\author{\IEEEauthorblockN{Nikos Fotiou, Vasilios A. Siris, George C. Polyzos}
\IEEEauthorblockA{Mobile Multimedia Laboratory\\
Department of Informatics, School of Information Sciences and Technology\\
Athens University of Economics and Business, Greece\\
\{fotiou,vsiris,polyzos\}@aueb.gr}
}

\maketitle

\begin{abstract}
We propose a capability-based access control technique for sharing Web resources, based on Verifiable 
Credentials (VCs) and OAuth~2.0. VCs are a secure means for expressing claims about a subject. 
Although VCs are ideal for encoding capabilities, the lack of standards for exchanging and using 
VCs impedes their adoption and limits their interoperability. We mitigate this problem by integrating 
VCs into the OAuth~2.0 authorization flow. To this end, we propose a new form of OAuth~2.0 access 
token based on VCs. Our approach leverages JSON Web Tokens (JWT) to encode VCs and takes advantage 
of JWT-based mechanisms for proving VC possession. Our solution not only requires minimum changes 
to existing OAuth~2.0 code bases, but it also removes some of the complexity of verifying VC claims
by relying on JSON Web Signatures: a simple, standardized, and well supported signature format. 
Additionally, we fill the gap of VC generation processes by defining a new protocol that leverages 
the OAuth~2.0 ``client credentials'' grant.   
\end{abstract}


\section{Introduction}
With the advent of Cloud-based services, controlled sharing of resources over the Web has become
fundamental. Nevertheless, existing sharing techniques are based on rudimentary mechanisms, 
such as mere bearer tokens, or hard to guess URLs; consequently, there is a need for more flexible 
and decentralized solutions~\cite{Bir2014}. In this paper, we focus on resources
hosted in multi-tenant Web servers (e.g., a Cloud-based storage) and we examine the case
where each tenant is an organization that wishes to control the access rights of its members in
an efficient and self-sovereign manner. In order to achieve our goal, we build a capabilities-based
access control solution that leverages OAuth 2.0 and Verifiable Credentials.   

OAuth 2.0~\cite{oauth} is a popular protocol used for authorizing $3^{rd}$ party \emph{clients}
to access resources stored in a \emph{resource server}. In OAuth 2.0 the authorization decisions
are made by a trusted entity referred to as the \emph{authorization server}. In many deployments
an authorization server and a resource server belong to the same administrative entity, hence it
is sufficient to encode authorization decisions in opaque \emph{bearer} tokens and enforce
them using simple \emph{Access Control Lists} (ACL). For example, an authorization server can be a social
network and a resource can be the profile of a specific user. Using OAuth 2.0 a Web application may receive
a token that includes a random string: this string is stored in an ACL of the social network
and is associated with an `access' right to a specific profile. Nevertheless, in scenarios where
the resource server is independent of the authorization server, using an ACL is cumbersome or
even impossible to use (e.g., if a resource is not connected to the Internet). A solution for this case
is to encode access control decisions in \emph{capability tokens}, i.e., tokens that describe
the \emph{access rights} of their holders. In this paper we use Verifiable Credential to build
such capability tokens. 

A Verifiable Credential (VC) provides a cryptographically secure, privacy preserving, and machine-verifiable means 
for expressing real-world credentials in the cyber world. In contrast to standard public-key based certificates that 
provide a binary identification, i.e., either the whole identity of the subject 
is disclosed, or nothing, VCs can be used 
for verifying certain attributes of a subject~\cite{Bau2008}. From a high-level
perspective, a VC allows an issuer to attest some claims about a subject, then,
the subject can prove ownership of these claims to a verifier. However,
the procedures for requesting and using VCs are not yet standardized. In this paper,
we fill this gap by leveraging the authorization flows of OAuth 2.0. 

Our solution allows a resource owner to use an OAuth 2.0 authorization server
for specifying the capabilities that $3^{rd}$ party clients have over the owner's
resources. Then, these clients can request from the authorization server an
``access token,'' which is granted in the form of a VC that includes the client's
capabilities. Eventually, this
VC can be used for gaining access to a protected resource. Using a VC as an access
token has several advantages: a VC has usually longer lifetime since it can
be used only by its holder (as opposed to a mere bearer token) and soon it will
be possible to store them in secure ``wallets,'' VCs can be combined, the VC data
model supports more efficient and privacy preserving revocation mechanisms, and
finally it is possible to agree on different ``types'' of VCs each of which can
use pre-defined claims with well-known semantics.  
With this paper we make the following contributions:
\begin{itemize}
    \item We define an OAuth 2.0-based protocol for issuing VCs.
    \item We integrate VCs into JSON Web Tokens, which enables us to use well-defined
    standards for expressing user identities and digital signatures.
    \item We build on ongoing standardization efforts for demonstrating secret key proof-of-possession, which are
    designed for OAuth 2.0 workflows and achieve their goal with a single message exchanged between
    the key holder and the verifier. 
    \item We design a Cloud storage system and we show that our solution is lightweight
    and can be easily integrated in existing systems. 
\end{itemize}

The remainder of this paper is organized as follows. In Section 2 we introduce the technologies
that are used as building blocks in our solution. In Section 3 we detail our design. We present
the implementation and evaluation of our solution in Section 4. We discuss related work in 
Section 5, and we conclude our paper in Section 6. 

\section{Background}
\subsection{JSON Web Keys, JSON Web Tokens, JSON Web Signatures}
A JSON Web Key (JWK)~\cite{rfc7517} is JSON data structure that represents a cryptographic key.
The following listing is the JWK-based representation of an Ed22519 public key~\cite{Ber2012}.
\begin{minipage}{\linewidth}
\begin{lstlisting} [caption={An example of JWK}]
{
  ``crv": ``Ed25519",
  ``kty": ``OKP",
  ``x": ``2Qt...zYAN0hxXfI0YbBtSG4eYY"
}
\end{lstlisting} 
\end{minipage}

A JSON Web Token (JWT)~\cite{Jon2015} is a compact, URL-safe means of representing ``claims''
as a JSON object. IANA's ``JSON Web Token Claims'' registry includes a list of ``registered''
claim names. Some of the standard claims we are using in our examples are:
\begin{itemize}
  \item \texttt{jti}: a token identifier.
  \item \texttt{iss}: a token issuer identifier.
  \item \texttt{iat}: a timestamp indicating when the token was issued.
  \item \texttt{exp}: a timestamp indicating the token's expiration time.
\end{itemize}

A JWT can be encoded in a JSON Web Signature (JWS)~\cite{rfc7515}. 
JWS represents content secured with digital signatures (or Message Authentication Codes) using JSON-based
data structures. A JWS object consists of a header (also known as the JOSE header), the payload, and
the signature. The JWS header includes information about the type of the payload and the signature
algorithm. The following listing is an example of a JWS containing a JWT in its payload (the signature
part is omitted).
\begin{minipage}{\linewidth}
\begin{lstlisting} [caption={An example of a JWT encoded in a JWS (signature omitted)}]
{
  ``typ": ``jwt",
  ``alg": ``EdDSA"
}.
{
  ``iat": 1617577068,
  ``iss": ``https://mm.aueb.gr",
  ``jti": "0xf94f7328e2bb70f575"
}
\end{lstlisting} 
\end{minipage}

Lines 1-4 are the JWS header, which indicates that the payload is a JWT and the signing
algorithm is EdDSA. Lines 5-9 are the JWT payload.

\subsection{OAuth 2.0 and the client credentials grant}
\label{sec:oauth}
\begin{figure}
    \centering
    \includegraphics[width=0.9\linewidth]{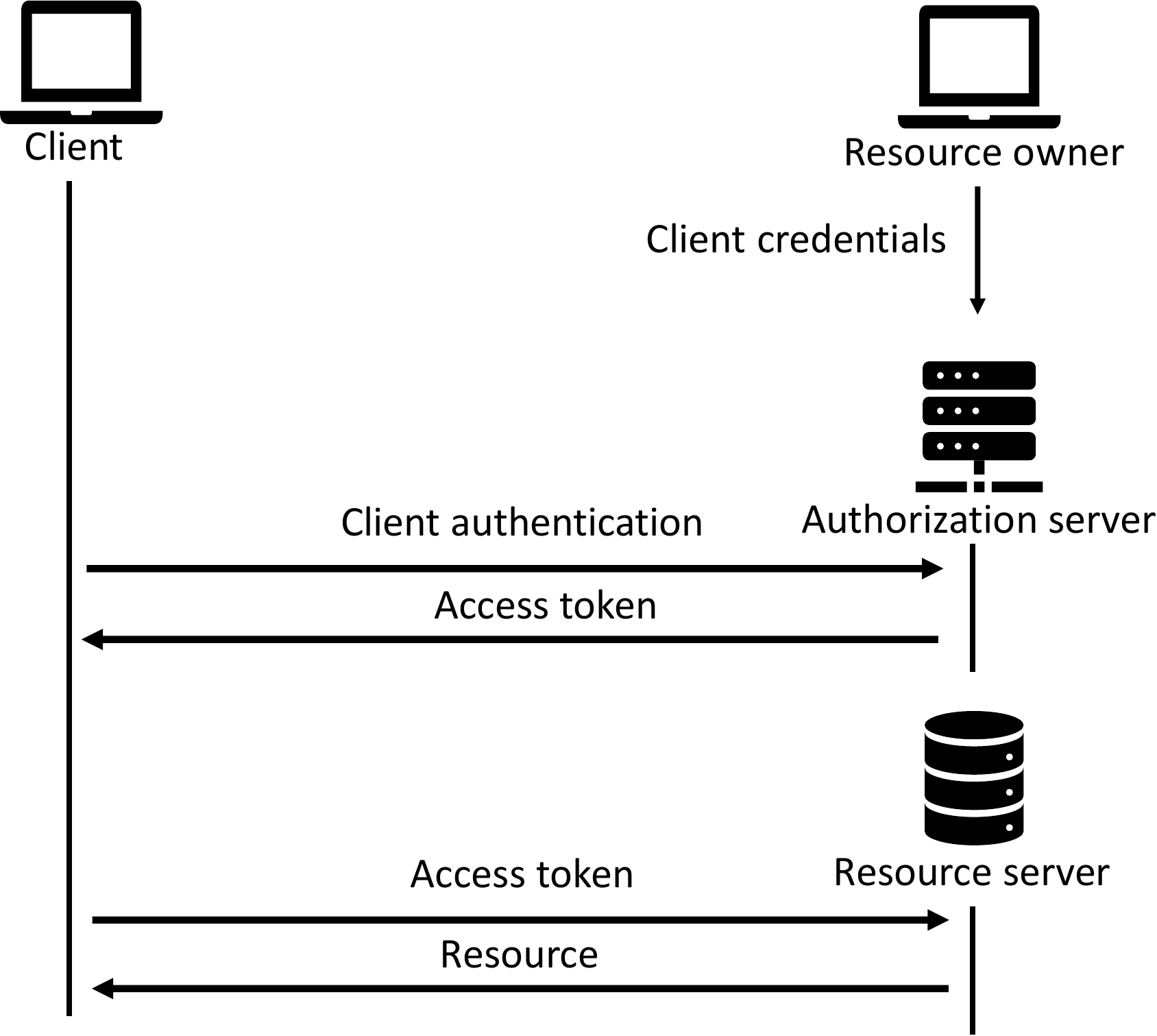}
    \caption {OAuth~2.0 interactions using a client credentials claim.}
    \label{fig:oauth}
\end{figure}

An OAuth 2.0 architecture is composed of the following entities.
A \emph{resource server} that hosts a protected resource owned by a \emph{resource owner}, 
a \emph{client} wishing to access that resource, and an \emph{authorization server} responsible 
for generating \emph{access tokens}. Access tokens are granted to clients authorized by the resource owner:
client authorization is proven using an \emph{authorization grant}. In our system we are using the
`client credentials' grant. As it can be seen from Fig.~\ref{fig:oauth}, when this type of
grant is used, a resource owner configures the authentication server with the credentials
of the authorized clients; a client \emph{authenticates} to the authorization server
and receives an \emph{access token}, then it uses the access token to access the protected
resource.

OAuth~2.0 deployments are free to choose their own types of access tokens. The solution presented in this paper 
uses as access tokens JWT-encoded VCs integrated in a JWS. The used tokens are enhanced 
with Proof-of-Possession (PoP) key semantics~\cite{rfc7800}.
In particular, they include a claim, named ``cnf'', whose value is a JWK representing the public key of the client
who owns the access token. Clients in our system must prove
ownership of the corresponding private key, otherwise the token is not accepted by the resource server.

\subsection{Verifiable Credentials}
\label{sec:bvc}
\begin{figure}
    \centering
    \includegraphics[width=0.9\linewidth]{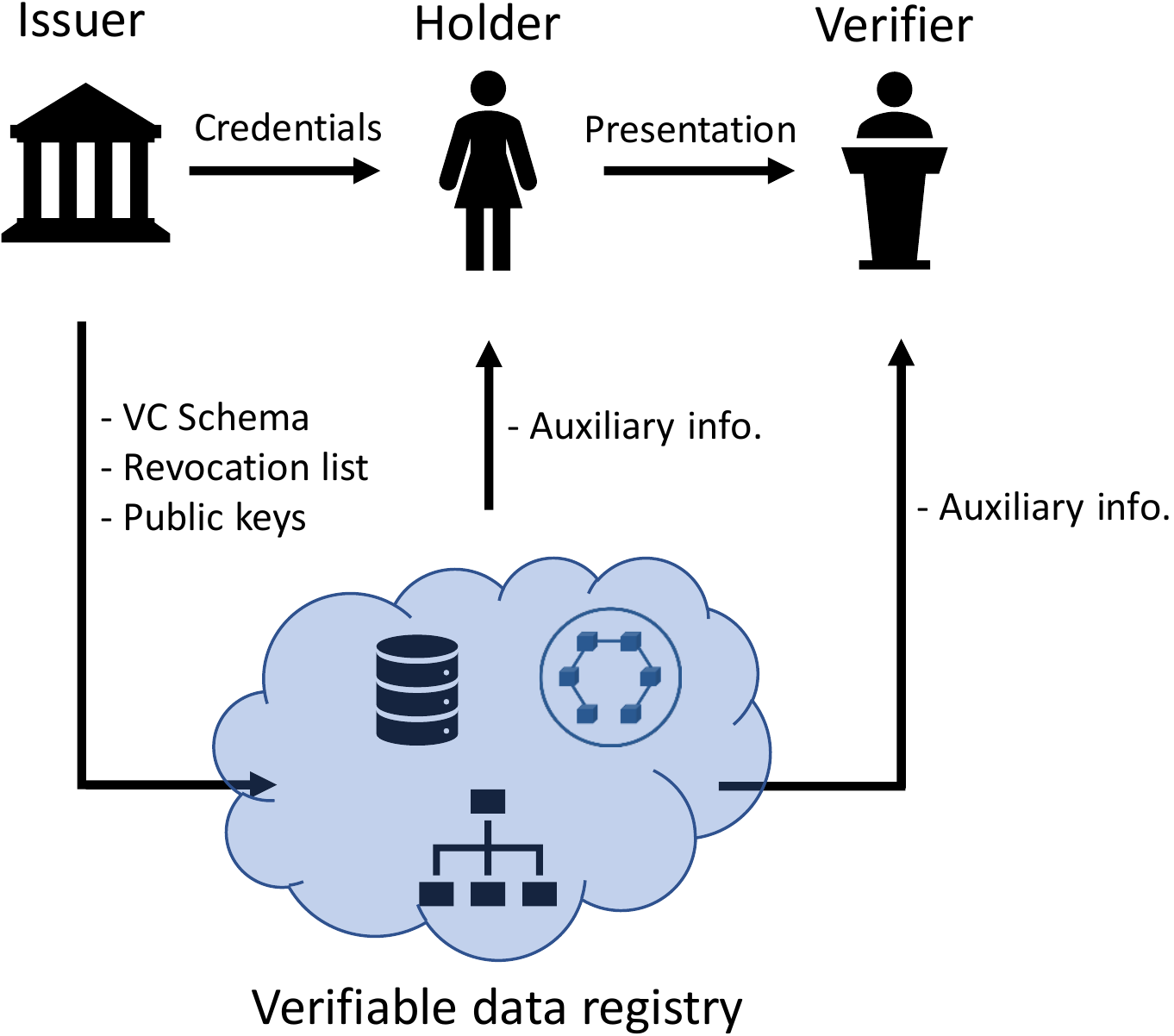}
    \caption {Verifiable Credential architecture.}
    \label{fig:vc}
\end{figure}

A Verifiable Credential (VC) architecture is usually composed (see also Fig.~\ref{fig:vc}) of an \emph{issuer},
a \emph{holder}, a \emph{verifier}, and a verifiable data \emph{registry}. These entities interact with each other
as follows. The issuer generates and signs a VC that attests some claims about a subject, then the VC is transmitted
to a holder. Usually a holder and a subject are the same entity;\footnote{In general, the holder could hold VCs for many subjects, e.g., a parent for underage children, or a person for her IoT devices.} this applies to our system
and in the following we will use these terms interchangeably. Holders can prove to a verifier
that they possess VCs with certain characteristics by generating a  \emph{Verifiable Presentation}~(VP).
In its simplest form, a VP can be a VC signed by the subject, but in the general case a VP may
include many VCs. All these operations are supported by a verifiable data registry. This registry
is used for tracking the status of a VC and it can be implemented as a centralized service or
in a distributed system, such as a blockchain. In our system the functionality of the registry
is centrally provided by the issuer. 

The W3C VC data model~\cite{ver} specification defines a standard way to express VCs using JSON-LD~\cite{jsonld}. 
The following listing includes a VC expressed in W3C's VC data model.

\begin{minipage}{\linewidth}
    \begin{lstlisting} [caption={A sample VC},label={list:vc}]
    {
      ``@context": ``https://www.w3.org/.../v1",  
      ``id": ``http://example.edu/credentials/3732",
      ``type": [``VerifiableCredential", ``UserId''],
      ``issuer": ``https://example.edu/issuers/14",
      ``issuanceDate": ``2021-01-01T19:23:24Z",
      ``credentialSubject": {
          ``id": ``did:example:ebfeb121",
          ``Name": ``Alice''
        }
    }
    \end{lstlisting} 
  \end{minipage}

Line 2 defines a \emph{context} for the VC. A context is a pointer to a VC description file
that defines various VC \emph{types}. Line 4 declares the type of the VC. VC types are a powerful
concept since different communities can come up and agree on certain VC types that specify
claims with well-understood semantics. For example, currently there are working groups
preparing such VC types that will represent academic records, travel documents,
medical history, and others. 

Line 3 defines an identifier for the VC which is used for tracking its status. Line 5
includes an identifier of the VC issuer (in this example the identifier is a URL), and
line 6 states the VC creation time and date.

Line 8 includes an identifier of the credential subject. This identifier must be a URI
which, if dereferenced, ``results in a document containing machine-readable information about the id.''
This document usually includes a public key which is then used for verifying a Verifiable Presentation.
In this listing a Decentralized Identifier (DID)~\cite{did-spec} is used as subject $id$, which is a
common practice in most VC implementations. Nevertheless, the process of retrieving a public key
from the subject $id$ is not defined. Even the DID standard, which has reached the status 
of ``Release Candidate'' 
only recently, allows each DID \emph{method} to define its own way for retrieving the public key
that corresponds to a DID. For this reason, our solution considers an  alternative
representation that allows the declaration of the subject's public key directly in the VC.

A VC is associated with a cryptographic \emph{proof} that makes the credential tamper resistant (usually
a digital signature). W3C's VC data model defines two proof methods: wrapping the VC in a JWT and use
JWS, or use \emph{linked data proofs}~\cite{ldp}. Our solution uses the former, which is a standardized,
widely used process. Linked data proofs have some advantages, such as they allow integration of
semantics into the proofs, they support partial proofs and proofs generated by different public keys, and 
they can use Zero-Knowledge Proofs. On the other hand, linked data proofs are still a community
draft and they rely on a complex canonicalization process that may potentially impede adoption
and create security threats.

\section{Design}
\subsection{Entities and Definitions}
\label{sec:def}
We now describe the entities of our system using, when possible, the terminology
of OAuth 2.0. Our system is composed of \emph{clients} belonging to different \emph{organizations}.
Each organization is the \emph{owner} of a resource, stored in a \emph{resource server} (RS)
shared among multiple organizations.
Furthermore, each organization maintains an \emph{Authorization Server} (AS).
The goal of our system is to allow owners to give permission to clients
to access some of the protected resources. Such permissions are expressed in the
form of \emph{access tokens}. Each AS has a token generation endpoint, which can be invoked by clients in order to
obtain an access token. 

\emph{Clients} are identified by
a public key and in the following we refer to this key as $Pub_{client}$. Similarly,
an AS is identified by a URL, $URL_{AS}$ and owns
a public key referred to as $Pub_{AS}$. In the following we denote the ``\emph{Capability} to perform an \emph{operation} 
to a \emph{resource}'' by $C_{operation \rightarrow resource}$ and we say that a client has  
the right to perform an operation $o_x$ in a resource $r_y$ if it can prove the possession
of a VC that includes $C_{o_x \rightarrow r_y}$.

Each AS maintains an \emph{access table} per RS that contains policies 
of the form $Pub_{client} \rightarrow [C_1, C_2, ... C_n]$, i.e., policies 
that map the public key of a client to a list of capabilities.\footnote{We recognize
that keeping a separate access table per RS is unoptimized, however we adopt
this approach for simplicity. The format of the access table does
not affect the overall design of our system.} Similarly, each RS maintains
a \emph{resource table} that maps a resource identifier to $URL_{AS}$ and to the corresponding
$Pub_{AS}$. Therefore, each entry in the resource table indicates the AS responsible
for managing access to that resource, as well as its key.  
Finally, we assume that each RS is 
supporting a ``well-known'' \emph{credential definition}.

Our design relies on the
security of OAuth 2.0 and imposes the same security requirements~\cite{oauth2best}.

\subsection{Demonstrating proof-of-possession of a key}
There are many methods for proving the possession of a key. Our system uses ``OAuth 2.0 Demonstration 
of Proof-of-Possession at the Application Layer'' (DPoP)~\cite{dpop}. Although DPoP is still an IETF
draft it is receiving increased attention and it is being actively developed. Nevertheless, other
PoP mechanisms can be considered.

DPoP has been designed for HTTP communication and achieves PoP in a single message. In particular,
with DPoP key holders can include in their HTTP requests a header referred to as DPoP proof. A
DPoP proof is a JWS
signed using the key of the user. The JWS header includes a \emph{type} field, which 
is always set to ``dpop+jwt'', a digital signature
\emph{algorithm}, and a JWK \emph{public key}. The latter public key is 
used for verifying the signature of a DPoP.
The JWS payload includes at least a unique \emph{identifier} (which can be a sufficiently large
random number), the HTTP \emph{method} of the request, the HTTP \emph{URI} of the request, and the \emph{time}
when the request was created. Listing~\ref{list:vc} is an example of a DPoP proof used in our system.
Lines 1-9 are the JWS header and lines 10-15 are the JWS payload. Lines 4-7 include the key that can
be used for verifying the digital signature of the DPoP proof (not shown in the listing). Finally, this 
DPoP proof is used for a HTTP POST request to ``https://mm.aueb.gr/token''.

\begin{minipage}{\linewidth}
    \begin{lstlisting} [caption={Example of a DPoP proof--the digital signature is not shown.},label={list:dpop}]
{
  ``typ": ``dpop+jwt",
  ``alg": ``EdDSA",
  ``jwk": {
    ``crv": ``Ed25519",
    ``kty": ``OKP",
    ``x": ``2Qt...zYAN0hxXfI0YbBtSG4eYY"
  }
}.
{
  ``htm": ``POST",
  ``htu": ``https://mm.aueb.gr/token",
  ``iat": 1617548847,
  ``jti": ``0xd21a57f53c8...dae132f2"
}
    \end{lstlisting} 
  \end{minipage}

\subsection{Access token request}
With the access token request protocol, a client requests an access token from an AS. The request 
includes a DPoP proof with a signature that can be verified using $Pub_{client}$.
The AS verifies the DPoP proof as follows:
\begin{enumerate}
    \item It extracts $Pub_{client}$ from  the JWS header of the DPoP proof.
    \item It verifies the signature of the proof using $Pub_{client}$.
    \item It verifies that the proof payload includes the same HTTP method and URI as the client request.
    \item It verifies that the random number included in the proof payload has not been reused.
    \item It verifies that the proof is ``sufficiently fresh'' based on the 
    creation time included in the proof payload.
\end{enumerate} 

If the proof is valid, the AS locates $Pub_{client}$ in its access table and it constructs 
a VC (based on the RS credential definition) that includes all capabilities associated with $Pub_{client}$. 
Then it integrates the 
VC in a JSON Web Token (JWT) (as described in section 6.3.1 of~\cite{ver}). The produced
JWT includes (among others) the \texttt{iss} claim (which stands for \emph{issuer}) set to $URL_{AS}$,
and the \texttt{cnf} claim~\cite{rfc7800} set to the JWT-encoding of the $Pub_{client}$.
Finally, the AS encodes the generated JWT in a JWS signed using the private key of the AS.
An example of a generated JWS is included in Appendix A.
The JWS serialization using base64url is the access token and it is sent back to the client. 

\subsection{Resource request}
\label{sec:rec}
In the simplest case, a client requests to perform an operation on a
resource by providing a DPoP proof and the obtained access token.
We discuss requests that combine multiple VCs in section~\ref{sec:vp}. Upon receiving a request
the RS executes the following verification process:
\begin{enumerate}
  \item It extracts from the resource table the $URL_{AS}$ and the $Pub_{AS}$ that correspond to the requested resource.
  \item It deserializes the access token, it verifies that the \texttt{iss} claim is set to $URL_{AS}$ and it verifies
  its signature using $Pub_{AS}$.
  \item From the deserialized access token, it extracts the \texttt{cnf} claim and it verifies that it
  includes the same JWK as the DPoP proof.
  \item It verifies the signatures of the DPoP proof, as well as that it contains a valid HTTP URL and method,
  a unique identifier, and a recent creation time.
\end{enumerate}

The $3^{rd}$ step of this process verifies the binding between the VC and the
client that sent the request. If all verification steps are successful, the RS extracts
the capabilities of the client from the VC and verifies that they are sufficient for performing
the requested operation. The implementation of the latter verification is
application and policy specific and we do not discuss it in this paper.   

\subsection{Token revocation}
With our approach, the access token status can be determined using two methods: either by
requesting the status of the JWT from the AS using OAuth 2.0 token introspection~\cite{rfc7662},
or by checking in a revocation list the status of the included VC using the method in~\cite{revoc_list}.

\subsubsection{OAuth 2.0 token introspection}
OAuth 2.0 token introspection allows a RS to query a \emph{token introspection}
endpoint, implemented by the AS, in order to determine
the state of an access token, as well as ``meta-information'' about this token.
However, the means by which the RS discovers that
endpoint are not defined.
Using this mechanism the RS sends to the token introspection endpoint the access
token, and the token introspection endpoint responds with a JSON object that contains,
among others, a field named \texttt{active} and it indicates the status of the token.

Even though this is a standardized mechanism, it has several disadvantages. Firstly,
it creates communication overhead, since the RS must query for the status of each token
individually. Secondly, it impairs client privacy since the AS can infer when each client
uses its access token. Finally, as already noted, it is not obvious how an RS can learn
the token introspection endpoint.

\subsubsection{VC revocation status}
An alternative solution for verifying the status of the access token is to 
extract the included VC and use the revocation scheme described in~\cite{revoc_list}.
This scheme is privacy-preserving and efficient, and although it is still
a draft, it is based on well understood mechanisms that are used by other similar
systems (e.g. such a recent system is described in~\cite{Smi2020}.)

Based on this revocation scheme, the AS maintains a revocation list that concerns all VCs it has issued. 
This list is a simple bitstring and each credential is associated with a position in the list. 
Revoking a VC means setting the bit 
corresponding to the VC to 1. Furthermore, each generated VC includes a field named
`revocationListIndex' that specifies the position of the credential in the revocation list.
Upon receiving an access token that includes one or more VCs that support this revocation 
mechanism, the RS must download the revocation list from the AS and examine the status
of the VC(s). It should be noted that the AS does not learn any information about the
VC(s) the status of which the RS wants to learn. Furthermore, since the revocation list includes information about
multiple VCs, an RS can download it once and use it for multiple VCs used
``soon enough'' after the list has been downloaded. Finally,
a VC may include a URL that the RS can use in order to access the revocation list,
therefore, a revocation list can even be stored in a location different from $URL_{AS}$.
This is another advantage of this method, compared to OAuth 2.0 token introspection. 

\subsection{Combining multiple VCs}
\label{sec:vp}
An important property of VCs is that a VC holder can combine multiple VCs in a single
\emph{Verifiable Presentation} (VP). Therefore, in our system, a client may receive multiple
access tokens, that may have been obtained by different ASes, and combine them in a single access token. The
caveat of this procedure is that each individual access token must contain the same $Pub_{client}$
in the corresponding \texttt{cnf} field. The process for creating such a VP is the following.
Initially, the client creates a new JWT object. The \texttt{iss}uer field of this JWT is 
set to the sha-256 hash of $Pub_{client}$. Furthermore, the generated JWT includes a field
named \texttt{vp} that contains an array of all individual access tokens. Finally, the generated
JWT is encoded in a JWS signed by the client; the serialized JWS is the new access token.

The process of verifying an access token is modified for the RS as follows: it deserializes
the access token and it examines if it contains a \texttt{vp} field; if it contains one, it extracts
all individual access tokens and verifies them
using the process described in section~\ref{sec:rec}, then it verifies the signature of the
received access token using $Pub_{client}$. If all verifications are successful, the RS extracts
all capabilities from all VCs and uses them to verify if they are sufficient for authorizing
the requested operation. 

\subsection{Using DIDs as VC subject}
\begin{figure}
    \centering
    \includegraphics[width=0.9\linewidth]{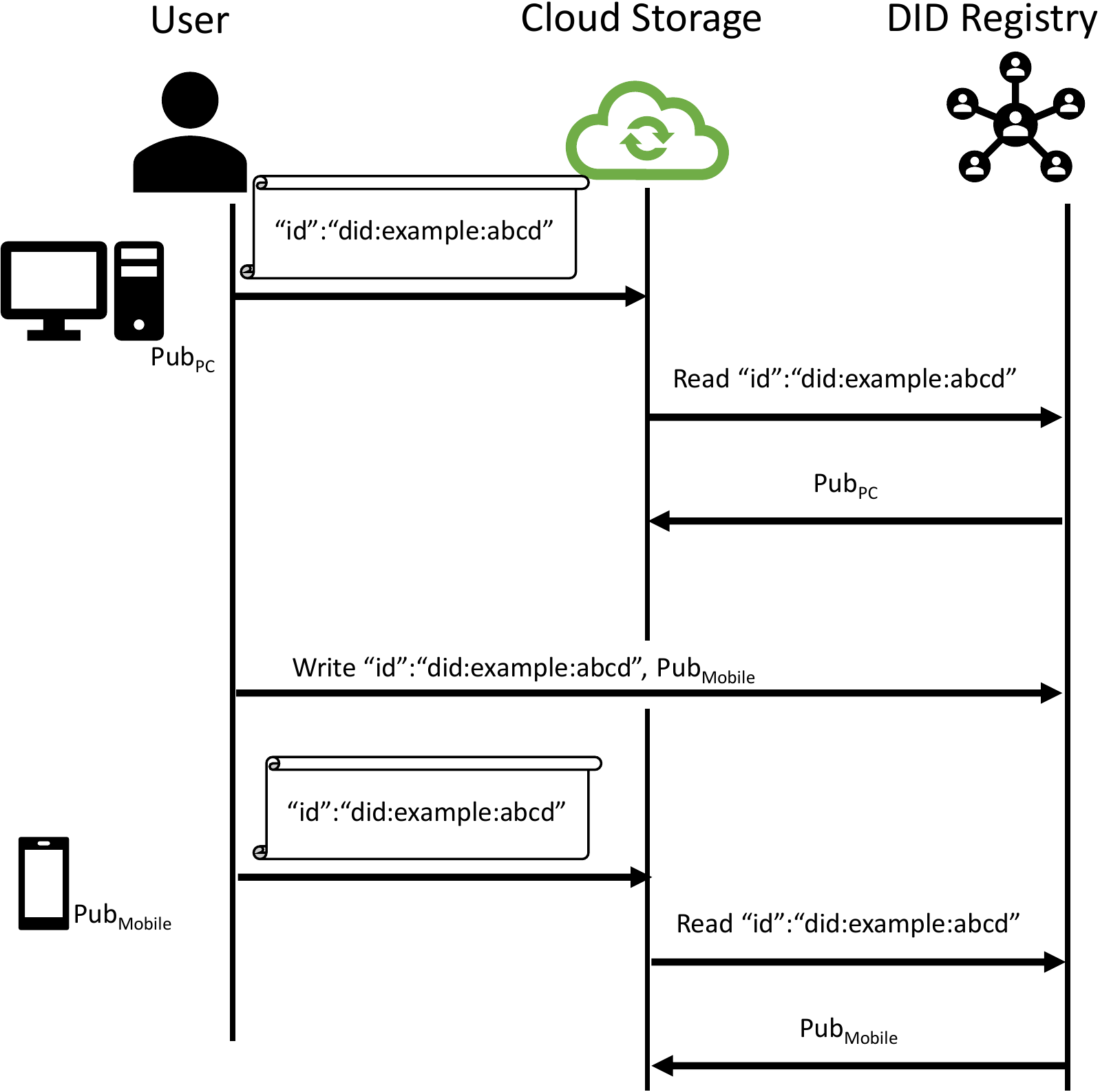}
    \caption {Key rotation using DIDs.}
    \label{fig:did}
\end{figure}

As discussed in section~\ref{sec:bvc}, most VC implementations use a Decentralized Identifier (DID)
to identify the credential subject, as opposed to our system that uses public keys. 
A DID has an important property that our system does not support: the public key associated with a
DID can be rotated. In order for an entity to learn the public key associated with a DID, it usually
has to query a DID \emph{registry}. 

Figure~\ref{fig:did} illustrates an example where key rotation 
is used. In this example a user uses as client her home PC. The PC has received an access token that
includes a DID. This token can be used with the process described so far, with the exception that
in order for the RS to validate it, it has to lookup in the registry the public key that corresponds
to the DID (this public key is used as the $Pub_{client}$). At some point the user decides to use
her ``travel phone'' for accessing the RS. She communicates with the registry and she updates
her DID to map to a public key stored in her phone. Now, when she repeats the resource access request,
using the same access token, the RS will learn a different $Pub_{client}$.

\section{Implementation and Evaluation}
We have implemented a proof-of-concept prototype of our solution using Python3. 
JWS functionality is provided by the JWCrypto library.\footnote{https://jwcrypto.readthedocs.io/en/latest/}
We are using Ed22519 public keys and the EdDSA digital signature~\cite{Ber2012}. 
\subsection{Cost of Authorization Primitives}
Our system entities are required to generate and validate JSON Web signatures.
In a desktop PC running Ubuntu 18.04, on an Intel i5 CPU, 
3.1Ghz with 2GB of RAM, JWS generation required 0.7ms and JWS verification requires
0.2ms.\footnote{Time intervals are measured using Python3's time module.}

Table I shows the number of signatures and verifications required per core
functionality of our system. 

\begin{table}
  \centering
  \caption{Cryptographic operations required per system operation.}
  \begin{tabular}{|c |c |} 

  \hline
  Functionality & JWS operations \\
  \hline
  Access token generation & 1 ver. + 1 sig. \\
  Resource request &  1 sig. \\
  Access token verification &  2 ver. \\
  \hline
 \end{tabular}
 \end{table}
\subsection{End-to-end system evaluation}
\begin{figure}
    \centering
    \includegraphics[width=0.9\linewidth]{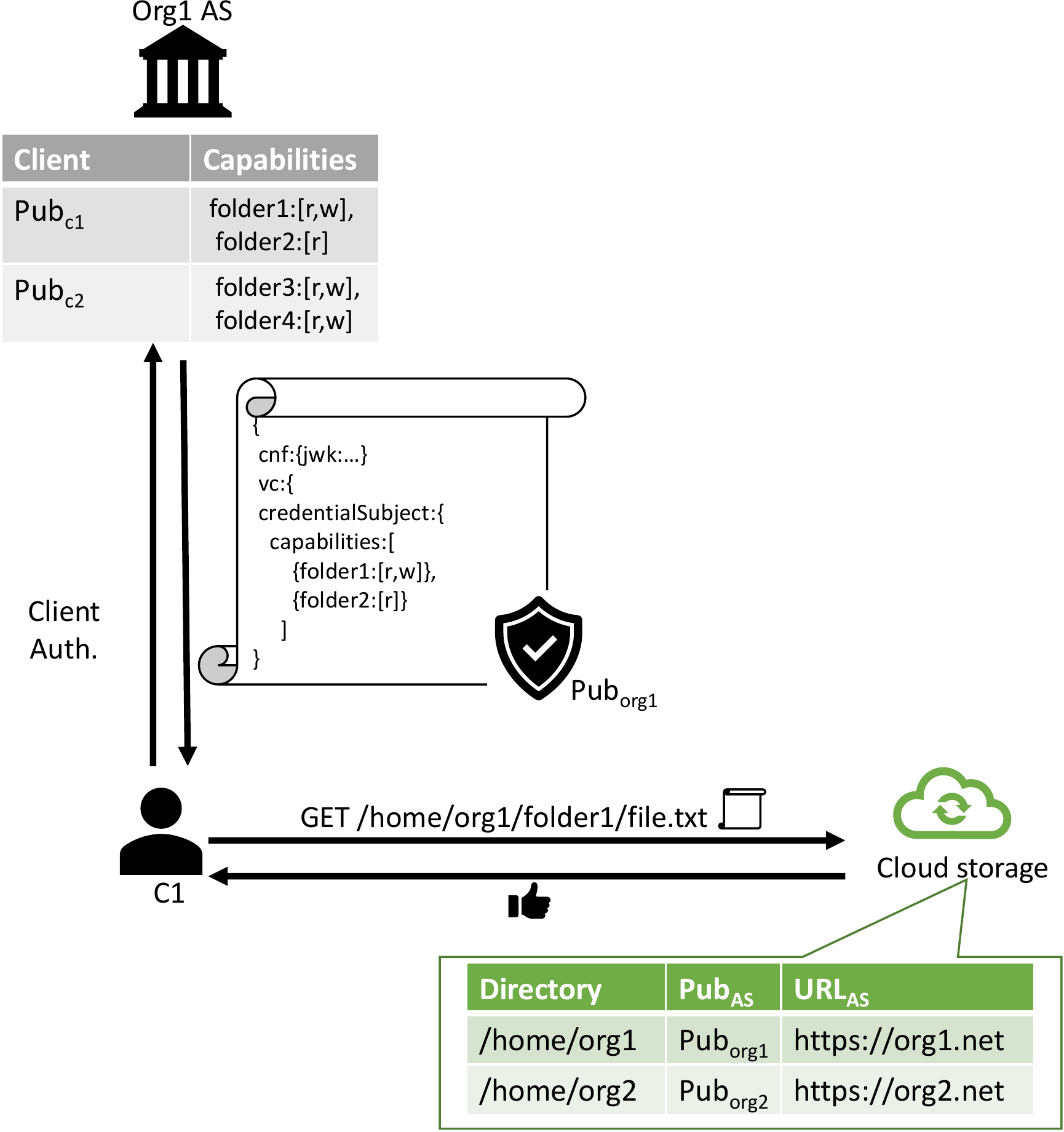}
    \caption {Our end-to-end system implementation.}
    \label{fig:impl}
\end{figure}

We now discuss a proof-of-concept implementation of our Cloud storage system described in
Section 1. Our system (illustrated in Figure~\ref{fig:impl}) includes a Cloud storage system
where multiple organizations store their files. Each organization implements an Authorization
Server (AS), which is responsible for assigning file access rights to the organization members. 

As discussed in Section~\ref{sec:def}, the Cloud storage 
maintains a \emph{resource table}. In our system the table contains entries that map a location in the file system
to the public keys of the AS that is responsible for managing access to that location. For example, 
in the instance of the system illustrated in Figure~\ref{fig:impl}, the AS that uses the public key
$Pub{org1}$ is responsible for managing access to the \texttt{/home/org1} directory.

Additionally, the Cloud storage has created a VC type definition that describes the format
that the VC claims should have in order to be accepted. In this example a VC should include
claims of the following form:

\begin{minipage}{\linewidth}
    \begin{lstlisting} 
{
  ``capabilities": [
      {<path> : [<access rights (r,w,d)>]}
  ]
}
    \end{lstlisting} 
  \end{minipage}

Therefore, a VC should include a key with name ``capabilities'' and its value will be a list
of paths and the corresponding read, write, delete access rights the subject has on this path.

Each AS maintains an access table with entries that map the public key of a client to the corresponding
access rights. In the example of Figure~\ref{fig:impl} the AS of org1 has an access table that includes
the public keys of two clients; the first client has read and write access to folder~1, and read access
to folder~2. Similarly, the second client has read and write access to folders~3 and 4. 

Supposedly, client $C1$ wishes to interact with the Cloud storage. He executes the \emph{access
token request} protocol and receives an access token. Then, he performs a
\emph{resource access request}. The \emph{access token request}
is an HTTPS POST request to the AS token generation endpoint and it includes in its body, in
addition to the fields required by OAuth 2.0, a DPoP proof encoded using base64url. The size
of a base64url encoded DPoP proof generated using EdDSA is $\sim430$ bytes (the size depends on the length
of the AS URL, which must be included in the DPoP proof payload). 
The AS responds with the signed access token also encoded in base64url.
The size of the access token depends on the number of the included claims. E.g., the
size of the access token that corresponds to the JWS included in Appendix~A is 719 bytes.
The resource access request is an HTTPS GET request that includes the \emph{Authorization} 
HTTP header, containing the access token, and an additional HTTP header named
\emph{DPoP}, containing the base64url encoded DPoP proof.

The Cloud storage performs the following validation operations.
\begin{itemize}
    \item It retrieves the public key of the AS responsible for the path \texttt{/home/org1}.
    \item It verifies the validity of the VC and the DPoP using the protocol of Section~\ref{sec:rec}.
    \item It examines if the capabilities included in the VC give to the client read access to the requested file.
\end{itemize}

\subsection{Solution properties and overall evaluation}

Our system has the following properties:

\textbf{It provides higher availability of the access control functionality.} Once the 
client receives the access token with the appropriate VC
it can interact with the Cloud storage even if the AS is not available. Moreover, and since a VC
has longer lifetime compare to simple bearer tokens, our system is more resilient to AS failures. 
Similarly, the Cloud storage can evaluate the validity of the access token without interacting with the AS with 
the caveat that if the revocation service is implemented by the AS, and the AS cannot be accessed,
the Cloud storage will not be able to 
determine the status of the access token.

\textbf{It improves client privacy.} Since a VC has a long lifetime, a client does not need to interact
with the AS whenever she wishes to access some files in the Cloud storage. Therefore, once the client
obtains the access token she can interact with the Cloud storage without notifying the AS. This privacy
feature is affected however by the used revocation protocol. In particular, with
``OAuth  2.0  token  introspection'' the Cloud storage sends to the AS the access token it received from the
client, therefore, the AS learns some information about the RSs with which a client interacts;
the granularity of this information depends on how often a RS needs to verify the status
of an access token. 

\textbf{It facilitates data availability.} An organization can store its data in multiple Cloud providers, 
thus increasing data availability, while using the same VC for allowing access to the same data from any 
of the Cloud providers. Thus, an organization can uniformly control user access to its data independently 
of the Cloud provider where the data is stored.

\textbf{It facilitates data portability.} In the scenario we described, the Cloud provider created
its own VC definition. But a powerful property of VCs is that various communities can agree on
common VC definitions. Therefore, in case there is a VC used for ``accessing Cloud storage''
an organization can migrate its data from one Cloud provider to another without needing to 
re-issue VCs to the users. This is particularly relevant for Edge computing and networking, 
where Edge services may be provided my many different ISPs or network providers.

\textbf{It facilitates multi-tenancy.} The Cloud storage provider can accommodate multiple organizations
and perform fine-grained access control decisions without having access to the user management systems
of the organizations, and without needing to understand the various ``roles,'' or ``structures,'' or
``policies'' each organization uses.

\textbf{It achieves self-sovereign access management.} Each organization can independently and without
having to interact with the Cloud storage provider manage which users can access each file. 

\section{Related work}
The lack of protocols for issuing and using VCs is a known problem and 
there are efforts that try to mitigate it.

Self-Issued OpenID Connect Provider DID Profile (SIOP DID)~\cite{siop} is an ongoing effort
of the Decentralized Identity Foundation\footnote{https://identity.foundation/} that builds on
OpenID connect~\cite{openid}. In particular, it leverages a feature of OpenID Connect that allows
users to act as OpenID providers and ``self-issue'' ID tokens (these are JWTs that include attributes
about a user). Therefore, with SIOP DID a user can issue a DID  (or VC) by herself. In our solution
issuers are decoupled from users. Furthermore, SIOP DID is only focusing on the ``last-mile,'' i.e.,
how a holder sends a VC to a verifier. Our solution instead considers how a holder will receive a VC, as well
as how she will prove legitimate possession.

Chadwick et al.~\cite{Cha2019} propose a solution for the lifecycle management of DIDs and VCs using
the FIDO Universal Authentication Framework (UAF). In particular they extend the \emph{user
registration} process of UAF allowing an ``identity provider'' (IdP) to install a VC to a
user device; this VC will be bound to the user identity. Furthermore, they extend 
FIDO's authentication protocol
to support user authorization as well. This is achieved using a handshake protocol with which
the ``service provider'' sends to the user its authorization policy, and the FIDO device
constructs and signs the appropriate verifiable presentation, which is then transmitted to
the service provider. The solution of~\cite{Cha2019} has similar properties to our system
and achieves similar goals. Nevertheless, our approach considers an additional entity,
the resource owner, who assigns capabilities to each client. Furthermore, the main
focus of FIDO UAF is user authentication,
which would require significant changes to be achieved.

Lagutin et al.~\cite{Lag2019} also integrate VCs and DIDs in OAuth~2.0.
However their solution is focused on IoT devices and they consider ACE-OAuth~\cite{ace}, i.e.,
a lightweight version of OAuth~2.0 that can be used in constrained devices over the CoAP
protocol. Furthermore, the solution proposed in~\cite{Lag2019} uses VCs and DIDs as authentication
grants that are used by the clients to obtain an access token from the AS. Our solution uses
the reverse approach: clients exchange an authentication grant to receive a VC as an access token. 

Apart from the efforts for integrating VCs (and DIDs) in existing protocols, there are
``clean slate'' approaches. For example DIDComm~\cite{didcomm} aims at building
a secure communication protocol based on DIDs. Presentation Exchange~\cite{difpre}
builds a protocol for allowing VC verifiers to request VPs from a holder. Similarly,
Hyperledger Aries~\cite{aries} develops a set of protocols, which can be 
used with Hyperledger's VC system. The credential handler API~\cite{chapi} defines 
an API that can be used by a website to request a user's credentials
and to help browsers correctly store user credentials for future use.
All these solutions however, not only require
re-implementing stakeholder software and lack interoperability with existing authorization standards,
but they are also focusing on a specific part of a VC lifecycle, e.g., Holder-Verifier communication,
or Holder-Issuer communications. 

In addition to a focus on the Web, there are efforts that try to integrate DIDs and VCs
in official identification systems. For example, Munoz~\cite{Mun2019} discusses how
DIDs and VCs can be integrated into eIDAS.

Our system leverages VCs for expressing capabilities because VCs are a standardized and well
understood technique. Nevertheless, capabilities-based systems consider two concepts that 
cannot be implemented using VCs in a straightforward manner: \emph{delegation} and \emph{attenuation}.
With delegation, a capability holder can transfer his capability to another entity, whereas
with attenuation he can confine a capability before delegating it. Consider for example 
our Cloud storage scenario; supposedly a client owns an access token that gives her access
to a set of files: using delegation and attenuation she could delegate her tokens to another
client limiting at the same time access only to a subset of the files allowed by the original
token. Two related approaches that can be used instead of
VCs in a system similar to ours are Macaroons~\cite{Bir2014} and 
Authorization Capabilities for Linked Data (ZCAP-LD)~\cite{zcap}.

\section{Conclusions}
In this paper we proposed a capability-based access control solution for
securing resources 
in multi-tenant resource servers. In order
to represent user capabilities we used the emerging standard of
Verifiable Credentials (VC). Furthermore, we leveraged OAuth 2.0 flows
for requesting and using VCs. Our solution used JSON Web Tokens (JWT) and JSON Web Signatures (JWS)
to encode VCs. JWT and JWS are well supported protocols and there 
already exists a very large code base. For this reason we believe
that our solution can be integrated in many existing systems.

One of the design choices we made was to not use Decentralized Identifiers (DIDs)
for identifying credentials of ``subjects,'' instead we used public keys. Although
we are missing some of the advantages of DIDs, such as key rotation, our solution
is easier to implement, and it is faster (since it does not require an additional
resolution step for mapping a DID to a public key). Nevertheless, all
of our constructions can be easily adapted to use DIDs instead of public keys.

OAuth 2.0 is the de-facto authorization protocol and it is used by many
systems. However, the vast majority of these systems assume short-lived
bearer tokens. As a result there are no browser-based ``token storage
systems,'' the notion of transferring one token from 
one device to another does not exist, and there are not efficient token revocation
mechanisms. On the other hand, VCs are supposed to have long lifetime,
and efficient mechanisms for storing and transferring them are being designed.
For this reason we believe that using VCs as access tokens, in the long run
will also result in better OAuth~2.0 related protocols.  

\bibliographystyle{IEEEtran}
\bibliography{IEEEabrv,references}
\section*{Appendix}
\subsection{Example of a JWS generated by an AS}
The following listing is an example of a JWS the serialization of which is used in
our system as an access token. Lines 1-4 are the JWS header, which declare that
the JWS payload is a JWT and the JWS is signed using EdDSA. Line~6 is the token
identifier, line~7 the issuer, and lines~8-9 the creation and expiration time.
Lines~10-16 are the client public key, encoded using JWK, and lines~17-34 are the
actual VC. Lines~23-27 provide information for tracking the revocation status of
the VC.

\begin{minipage}{\linewidth}
\begin{lstlisting}[caption={Example of a JWS generated by an AS}]
{
  ``typ": ``jwt",
  ``alg": ``EdDSA"
}.
{
  ``jti": ``https://mm.aueb.gr/credentials/1",
  ``iss": ``https://mm.aueb.gr/as",
  ``iat": 1617559370,
  ``exp": 1618423370,
  ``cnf": {
    ``jwk": {
      ``kty": ``OKP",
      ``crv": ``Ed25519",
      ``x": ``THpyF5W128...h5D4nb50qUU"
      }
   },
  ``vc": {
    ``@context": [
      ``https://www.w3.org/2018/credentials/v1",
      ``https://mm.aueb.gr/contexts/capabilities/v1"
    ],
    ``type": [``VerifiableCredential",``capabilities"],
    ``credentialStatus": {
      ``type": "RevocationList2020Status",
      ``revocationListIndex": "94567",
      ``revocationListCredential": "https://aueb.gr/rl"
  },
    ``credentialSubject": {
      ``capabilities": [
        { ``folder1": [``r",``w",``d" ] },
        { ``folder2": [``r"]}
       ]
    }
  }
}
\end{lstlisting} 
\end{minipage}

\end{document}